# Particles, fields and a canonical distance form

A. N. Grigorenko

*School of Physics and Astronomy, University of Manchester, M13 9PL, Manchester, UK*

We examine a notion of an elementary particle in classical physics and suggest that its existence requires non-trivial homotopy of space-time. We show that non-trivial homotopy may naturally arise for space-times in which metric relations are generated by a canonical distance form factorized by a Weyl field. Some consequences of the presence of a Weyl field are discussed.

# I. Introduction.

Classical physics describes motion of particles under an action of classical fields. Classical particles are usually assumed to be structurless material points. Classical fields are produced by charges that attract or repel each other. It is also conventionally assumed that elementary charges (or simply elementary particles) of classical physics are point-like and have vanishing spatial sizes. (This follows from the fact that classical solutions with charge distributed in some area of space are normally not stable. Hence unknown additional forces are needed to stabilise elementary particles if they were to occupy some finite region in space.) The classical picture therefore contains a space filled with delta-like charges and fields described by field potentials everywhere except the points of charge singularities [1].

It is also widely accepted that classical fields represent connections in a fibre bundle associated with a particle representation transforming under Lorentz and a local symmetry group of particle interactions [2]. There exists an asymmetry in dealing with particle representations and connections in classical physics: connections enter the scheme of classical physics (as field potentials) while particle representations (fibre co-ordinates on which these connection in an appropriate associated form act) often do not. For example, electromagnetic 4-potential (that represents connection in a space of complex particle representations) is an element of classical physics, while complex particle representations are not. As a result, fields lose their geometrical meaning in classical physics and appear to be ad-hoc assumptions of classical dynamics. In this light, it seems natural to eliminate the asymmetry and restore geometrical meaning of classical fields by adding an internal structure to a classical particle.

Recently we have discussed classical dynamics containing particle representations that transform under Lorentz and local symmetry groups of particle interactions [1]. We have assumed that every point of our space-time has its own copy of the additional particle coordinates (describing a state of the particle) and treated the space of classical physics as a fibre bundle. The local co-ordinates of the (associated) fibre bundle were $(x, |\phi\rangle)$, where $x$ are the usual space-time co-ordinates and $|\phi\rangle$ are the fibre co-ordinates (that transform under representations of the Lorentz and local symmetry groups).

We assumed that locally physics is simple and the fibre space of one point of space-time can be connected to that of an adjacent point by a linear connection. As a result, field potential plays a role of a connection in the world fibre bundle while a classical particle appears as a non-trivial state of the world fibre bundle described by a globally non-trivial connection. Since any non-trivial state of a world fibre bundle is accompanied by a non-trivial connection it implies that a classical particle is surrounded by fields and has some sort of singularity which is localised in space.

In Ref. 1 we "simplified" a classical particle to one point and assigned a particle representation vector to a point of its field singularity. The first conclusion of this approach was the fact that the conventional definition of geodesics should be modified when applied to classical particles with an internal structure. We managed to reformulate geodesics in terms of the parallel transport of the particle state vector $|\phi\rangle$ (instead of the parallel transfer of a tangent vector) under the price that the distance on a manifold, $ds$, should be determined by an eigenvalue of some operator-valued distance one-form $\hat{\theta}$: $\hat{\theta}(dx)|\phi\rangle = ds|\phi\rangle$ (instead of the conventional metric two-form). The new definition of geodesics is as follows: geodesic is a curve such that the parallel transport of the initial representation vector $|\phi\rangle$ to any point along the curve yields an eigenvector of the operator-valued distance form $\hat{\theta}(\dot{x})$ taken at this point (in accordance with the original definition where a geodesic is defined as a curve such that the parallel transport of a tangent vector along the curve gives a displacement on the manifold via the canonical forms $\theta^i$), see [1,3].

It turned out that the conventional metric two-form can be replaced by the linear operator one-form defining the same metrical relations. This linear operator was referred to as the canonical distance form (or simply the distance form) and has an analogy with the Finsler's metric. The action principle based on the distance form readily gives a description of classical particles with spin subject to Yang-Mills forces. The particle state $|\phi\rangle$ plays the role of the classical particle momentum in this description. We have shown that motion of spinor particles in this formulation of classical physics is affected by the space-time curvature.

In case of the four-dimensional Lorentz space-time (which is an area of low energy particles and fields) the canonical distance form can be written as $\hat{\theta}_s = \hat{\gamma}_a \theta^a$ (where $\hat{\gamma}_a$ are the Dirac matrices and $\theta^a$ are canonical forms of linear connection taken with respect to an orthogonal basis [1], $\theta^a = \theta^a{}_\mu dx^\mu$ and $\theta^a{}_\mu$ are vierbein fields) and the lowest bi-spinor representation consists of the direct sum of the left and right components. There exist very strong experimental indications that the left components of leptons (which has been used for measuring of distances) transform as SU(2) doublets ($L^a$) while the right components are SU(2) singlets ($R$). For example, the Standard Model requires the following local Yang-Mills symmetry group $G$=SU(3)×SU(2)$_L$×U(1)$_Y$. Many other theories of grand unification also suppose different actions of SU(2) group on left and right particles. In this case the distance form written as $\hat{\theta}_s = \hat{\theta}_{RL} \oplus \hat{\theta}_{LR}$ connects together spaces of different irreducible representations of SU(2) group ($L^\alpha$ is a SU(2) doublet and $R$ is a SU(2) singlet, $\alpha$=1,2) which is forbidden by Schur's lemma [4].

In order to remedy the situation by simplest means, we have introduced an additional scalar field $\varphi^\alpha$ which transforms as SU(2) doublet and glues spaces of left and right components of orthogonal representations. Then, the simplest canonical distance of our space-time is

$$\hat{\theta} = \hat{\theta}_{RL} \varphi^{\alpha\dagger} \oplus \varphi^\alpha \hat{\theta}_{LR} = \begin{pmatrix} 0 & \varphi^\alpha \hat{\theta}_{LR} \\ \hat{\theta}_{RL} \varphi^{\alpha\dagger} & 0 \end{pmatrix}. \tag{1}$$

Eigenvalues of the distance form (1) can be found as

$$\hat{\theta} \begin{pmatrix} L^\alpha \\ R \end{pmatrix} = d\lambda \begin{pmatrix} L^\alpha \\ R \end{pmatrix} \tag{2}$$

and yield the following length element $d\lambda$:

$$d\lambda^2 = |\varphi^\alpha|^2 \eta_{ab} \theta^a \theta^b = |\varphi^\alpha|^2 g_{ik} dx^i dx^k. \tag{3}$$

It is clear that the field $\varphi^\alpha$ scales the distance measured with the help of the particle that transforms as $\begin{pmatrix} L^\alpha \\ R \end{pmatrix}$. The idea to introduce a scaling factor into the length interval is not new and was proposed some years ago by Herman Weyl in a brilliant conjecture later transformed into the modern gauge theories [5]. The scalar doublet $\varphi^\alpha$ will be referred to as the Weyl field. This gives an action for a classical electron as:

$$S = \int \left[ \langle \phi | \hat{\theta} | \phi \rangle + \frac{i}{2} \chi \left( \langle \phi | \nabla \phi \rangle - \langle \nabla \phi | \phi \rangle \right) - d\lambda \left( \langle \phi | \phi \rangle - 1 \right) \right],$$

where $|\phi\rangle = \begin{pmatrix} L^\alpha \\ R \end{pmatrix}$, $\nabla = d + \begin{pmatrix} \hat{\omega}_{LL} & 0 \\ 0 & \hat{\omega}_{RR} \end{pmatrix} + \hat{\Omega} + \begin{pmatrix} \hat{\omega}_{LL}^{YM} & 0 \\ 0 & \hat{\omega}_{RR}^{YM} \end{pmatrix}$ ($\hat{\omega}_{LL}^{YM}$, $\hat{\omega}_{RR}^{YM}$ are the Yang-Mills connections for the left and right spinor components, $\hat{\Omega}$ is the Weyl form) and $\langle\phi| = \begin{pmatrix} \bar{R}q^{\alpha\dagger} & \bar{L}^\alpha q^\alpha \end{pmatrix}$ is the contragradient counterpart of $|\phi\rangle = \begin{pmatrix} L^\alpha \\ R \end{pmatrix}$ with $q^\alpha$ being an SU(2) vector [1].

The purpose of this paper is to provide two additional arguments in favour of the proposed distance form (1). Namely, we show that the distance (3) helps to solve problems of particle existence and singularities, discussed in Section II, and particle energy and divergences, discussed in Section III. We briefly discuss the Weyl field properties in Section IV. Finally, a conclusion is given.

**II. Particle existence and singularities.**

Created by the works of Weyl, Einstein, Cartan, Yang and Mills, gauge theories form the basis of the modern physics. They appeal to natural knowledge that locally physics is simple (there exists a local trivialisation of the world fibre bundle). Mathematically, they reflect the fact that any reasonable dynamics produces a flow which can be parallelised in appropriate coordinates (e.g., see the Darboux theorem [6]). Gauge theories describe behaviour of fields and motion of test particles extremely well. However, they lack one important ingredient, namely, a source of fields. A source of the field (a particle) cannot be described in a framework of a gauge theory in four-dimensional space-time. This follows from the following theorem:

*Any principal fibre bundle with a homotopically trivial base M (or the structure group G) is trivial.*

Mathematical details connected with this "triviality" theorem are given in Ref. 7, see Corollary 11.6. (Here we only note that *M* should be normal locally compact and such that any covering by open sets is reducible to a countable covering. A manifold is called homotopically trivial if it is contractible on itself to a point. A trivial fibre bundle is defined as the product bundle or, equivalently, as the one that allows a global cross section.) The manifold of our space-time is topologically identical to $R^4$ and is homotopically trivial. As a result, the triviality theorem forbids our space-time to have non-trivial fibre principal bundles which could be associated with particles. (Here by a trivial fibre bundle we imply one that allows a global cross section.) It is easy to show that in a trivial fibre bundle charges are distributed in the space (the total charge in a volume goes to zero when the volume goes to zero) and hence elementary particles would require additional forces to stabilize them. All experimental observations so far indicate that particles are non-trivial fibre bundles that do not permit a global cross section of space-time. Hence, we have to admit the presence of singularities inside our space-time in order to introduce classical particles. (In parenthesis we note that associated fibre bundles could be non-trivial even over homotopically simple bases, but these bundles would still lead to distributed charges and hence cannot be used to describe classical elementary particles. Also, we assume here that the field distribution is regular at infinity which excludes instanton-like or monopole-like solutions, where the limit of the field could be different for different wordlines approaching infinity.)

Indeed, by removing a point from $R^4$ ($R^4$ is topologically identical to space-time), we obtain a manifold $R^4 \backslash R^0$ which is topologically equivalent to the product $S^3 \times R^1$ and which allows non-trivial fibre bundles ($\pi_3(S^3 \times R^1) = \pi_3(S^3) = Z$, where $\pi_3$ is the homotopy group [7]). These fibre bundles could be associated with "photons" of the field because they would have only one singular point in space-time: a point of a "photon" creation or absorption. Analogously, by removing a singular line from $R^4$, we create a manifold $R^4 \backslash R^1$ topologically equivalent to $S^2 \times R^2$ which is also non-trivial ($\pi_2(S^2 \times R^2) = \pi_2(S^2) = Z$). Fibre bundles over this manifold could be related to a particle and the singularity line $R^1$ can be regarded as a particle world line. Finally, by removing a singular plane from $R^4$, we construct a manifold $R^4 \backslash R^2$ which is equivalent to $S^1 \times R^3$ with $\pi_1(S^1 \times R^3) = \pi_1(S^1) = Z$. Fibre bundles associated with this manifold could be linked to the Dirac monopole or vortices since they would have a singular line in three-dimensional space.

The triviality theorem is a generalisation of a well-known physical fact that the change density associated with an elementary particle is usually singular and that its charge is normally quantized. Hence an elementary particle cannot be described by a trivial fibre bundle that generates finite charge density. Physicists realised this problem a long time ago and various attempts have been made to develop a singularity-free theory of matter [8]. These attempts did not lead to a consistent and self-contained theory. As a result, several different approaches are now used to deal with singularities.

The most common is a positivistic approach which admits that something is wrong with a definition of an elementary particle but discards all difficulties. Physicist-positivist states that the main task of science is to predict results of measurements. Thus, scientists should not be interested in a detailed structure of nature as long as we can calculate every measured quantity. The theory of renormalization (developed by positivists) deals with infinities and singularities in exactly this vein. This is a consistent and successful approach shared by many.

Another approach (proposed by Kaluza and Klein [9]) is based on additional dimensions. This approach assumes that the base of our world fibre bundle is not equivalent to simple $R^4$ and introduces additional dimensions which are hiding from our observations. Then, the base of the world fibre bundle could be topologically non-trivial and hence non-trivial fibre bundles describing particles are possible. This attractive view has its advocates in a number of modern string theories. However, there is a difficulty connected with such an approach. Namely, using same arguments of covering homotopy one can prove that [7]:

*A fibre bundle P(M,G) with a homotopically simple base M produced by a reduction of a principal fibre bundle P(N,G) (M⊂N) is trivial.*

Thus, no matter how complex and non-trivial the fibre bundle is in the world with additional dimensions it will be trivial after a reduction to $R^4$ which is an area of low energy particles and fields. It implies that additional compactified dimensions should show themselves in observed space-time (in order to form a classical particle) for which we simply do not have enough experimental evidence at present.

A third common approach consists in ignoring problems connected with sources of fields on the basis that classical theory is not satisfactory anyway and quantum physics is needed for adequate description of our world. However, this approach just moves the problem of singularities deeper and deeper into quantum physics (from non-relativistic quantum mechanics to relativistic one, then to quantum theory of fields and then to a string theory) and finally leaves it without an answer at the point where we are not sure what are space-time, fields or particles. Also, this line of considerations fails for the Einstein theory of General Relativity, which can be formulated as a self-consistent classical gauge theory, where the problem of singularities attracted a lot of attention through the works of Penrose and Hawking [10].

The length element (3) provides an elegant way of solving the problem connected with particle existence and stability in classical physics. It is clear that any region where the scalar field $\varphi^\alpha$ is zero ($|\varphi^\alpha|=0$) is singular with undefined metrical relations. The base of the world fibre bundle is not simple $R^4$ in the presence of such regions and hence non-trivial fibre bundles over such base are possible. In accordance with the discussion above, a fibre bundle in which the Weyl field $\varphi^\alpha$ is zero at some point could be identified with a "photon" of the field, a fibre bundle which contains a region where the Weyl field equals zero along some line could be identified with a particle and a fibre bundle with $|\varphi^\alpha|=0$ over some plane in $R^4$ could be identified with a monopole or a vortex. We still need a homotopically non-trivial group in order to generate a non-trivial fibre bundle. The local symmetry group $G$=SU(3)×SU(2)$_L$×U(1)$_Y$ after retracting the remaining electromagnetic symmetry U$_{em}$(1) is good enough to ensure non-trivial particle-like principal fibre bundles because $\pi_2(G/U_{em}(1))=\pi_1(U_{em}(1))=Z$. These bundles even have a topological charge.

It is worth stressing that particles in this picture appear at places where the magnitude of the Weyl field goes to zero. This is in a stark contrast with standard Higgs-based models where the mass of elementary particles is produced by a non-zero value of the magnitude of Higgs field. A contribution of the Weyl field to particle energy is always non-zero which means that elementary particles which are described by a line where the Weyl field is zero should have non-zero masses. The close physical analogy to the proposed model of a particle is a vortex in type-II superconductors. The "universe" of superconductor is described by a wave-function of Cooper-pair condensate. This universe allows non-trivial fibre bundles with the structure group U(1) whenever the condensate density is zero in some region of superconductor. Due to symmetry, these fibre bundles are topologically stable when a region of vanishing condensate density is a line in $R^3$ (or a plane in $R^4$).

**III. Particle energy divergences.**

The distance form (1) and the length element (3) also ensure that the energy connected with particle singularities is finite. Indeed, according to (3) the volume element is proportional to $|\varphi^\alpha|^4$ and hence compensates an apparent divergence of energy of fields generated by the particle in the places of singularities with $|\varphi^\alpha| = 0$.

It is necessary to note that the tetrad defined by the distance form (1) is orthogonal but is not orthonormal. We can rewrite (1) as

$$\hat{\theta} = \tilde{\tilde{\theta}}_{RL} \frac{\varphi^{\alpha\dagger}}{|\varphi^\alpha|} \oplus \frac{\varphi^\alpha}{|\varphi^\alpha|} \tilde{\tilde{\theta}}_{LR} \qquad (4)$$

and the length element as

$$d\lambda^2 = \eta_{ab} \tilde{\theta}^a \tilde{\theta}^b, \qquad (5)$$

where $\tilde{\theta}^a(\varphi^\alpha)$ is an orthonormal tetrad that depends on the Weyl field. It is a canonical form of the length element of classical physics except for our knowledge that the tetrad is also defined by some scalar field. If we assume that this dependence is absent and the field $\varphi^\alpha$ is constant, we return to the case where particle fibre bundle are impossible and energies connected with "manufactured" particles are infinite.

There exists a good reason for moving the Weyl field $\varphi$ into a "geometry" part of the action and writing the length element in the form (5) conventional for classical physics. Let us consider a generic example of a scalar field $\varphi$ that defines the length element coupled to a gauge field of connection $A$ (we assume for simplicity that the contribution from fermionic fields can be neglected). The field action for this system could be written as

$$S = S_\varphi + S_A = \int \left[ \chi D\varphi^\dagger \wedge *(D\varphi) + V(\varphi) d\eta \right] + \frac{1}{4} \int \mathrm{tr} F \wedge *F, \qquad (6)$$

where $\chi$ is a constant, $V(\varphi)$ is a potential of the field $\varphi$, $d\eta$ is the volume element, $F = DA$ is the field strength (curvature) and the star denotes the Hodge operator. From (6) we get the following set of Maxwell equations:

$$\begin{aligned} D*F &= *j \\ D*j &= 0 \\ DA &= F \\ DF &= 0 \end{aligned}, \qquad (7)$$

where $\delta_A S_\varphi = \int \mathrm{tr}\delta A \wedge *j$ ($j$ is the current). Suppose that we know the solution $\varphi$ of the system (7). Introducing a decomposition $\varphi = T(g)\varphi_0 |\varphi|$ ($T(g)$ is an element of the local symmetry group and $\varphi_0$ is a fixed normalised vector), we find that the current associated with $\varphi$, is distributed in space since in general $|\varphi|$ is not constant.

We note, however, that there is no a clear way of separating a contribution from the Weyl field to an experimentally measured length interval. Also, this contribution could be different in different points of space-time reflecting different choice of units for the Weyl fields in different points. Hence, we have reasons to believe that the action for the Weyl field $\varphi$ should be scale invariant and allow local conformal symmetry. The arguments in favour of a conformal invariance of underlying physics have been already suggested by Weyl himself [5], see other works in review [11]. Let us, therefore, consider a generic conformal invariant action given by [11]

$$S = \int \left[ \chi D\varphi^\dagger \wedge *(D\varphi) - \frac{|\varphi|^2}{2} R^{ab} \wedge \eta_{ab} + \frac{\Lambda}{4} |\varphi|^4 \, d\eta \right] + \frac{1}{4} \int \mathrm{tr} F \wedge *F, \qquad (8)$$

where $R^{ab}$ is the Ricci tensor. The action for the Weyl field $\varphi$ contains the kinetic term $\int \chi D\varphi^\dagger \wedge *(D\varphi)$, the Penrose-Chernikov-Tagirov term $\int \frac{|\varphi|^2}{2} R^{ab} \wedge \eta_{ab}$ and the term proportional to the total volume of the space-time $\int \frac{\Lambda}{4} |\varphi|^4 \, d\eta$. In this conformal case, we can use a "geometry" trick and move a problem of $|\varphi|$ variation into the coordinate part of the action. Instead of

$$\delta_A S_\varphi = \int \mathrm{tr}\delta A \wedge *j \qquad (9)$$

we write

$$\delta_A S_\varphi = \int \mathrm{tr}\delta A \wedge \tilde{*}j, \qquad (10)$$

where the Hodge operator of the new metric relations is produced by the conformal group transformation of $\tilde{\varphi} = C\varphi$ with

$$C = \frac{|\varphi_0|}{|\varphi|}. \qquad (11)$$

Under this conformal transformation, the transformed Weyl field is given by $\tilde{\varphi} = T(g)\varphi_0$ whenever $|\varphi| \neq 0$. Hence the variation of this field in space outside the regions of $|\varphi| \neq 0$ can be removed locally by the gauge transformation $T(g)$ resulting in the zero current. In this case Maxwell equations well outside the regions of $|\varphi| = 0$ would have a simple solution $A = T^{-1}dT + \Omega + A_0$, where $A_0$ is the solution of the homogeneous equation $D*F = 0$ in $\mathrm{R}^4$ with exclusion of the points $|\varphi| = 0$ and $\Omega$ is the Weyl field that corresponds to the conformal group transformation of $\tilde{\varphi} = C\varphi$. This implies that particles and their gauge field are formed near the regions of $|\varphi| = 0$ in agreement with the fact that the base of fibre bundle is not topologically simple when the regions of $|\varphi| = 0$ are present. We arrive at almost a classical picture of "singular" particles with particle world lines being the lines of $|\varphi| = 0$. The "geometry" trick simplifies the field part of the action (6) considerably. Let $F_m$ be a part of the field strength produced by $m$-region of $|\varphi| = 0$ and $A_m$ is the corresponding part of the connection. Then we have (in new coordinates)

$$\int \mathrm{tr} F \wedge *F = \sum_{m,n} \int \mathrm{tr} F_m \wedge *F_n = \sum_{m,n} \int \mathrm{tr} DA_m \wedge *F_n = -\sum_{m,n} \int \mathrm{tr} A_m \wedge D*F_n =$$
$$-\sum_{m,n} \int \mathrm{tr} A_m \wedge *j_n = -\sum_{m} \mathrm{tr} A_m I_m - \sum_{\substack{m,n \\ m \neq n}} \mathrm{tr} A_m I_n \quad , \quad (12)$$

where $I_m$ is a linear current connected with $j_m$. (Integrations by parts and motion equations have been used.) The field part of the action is now represented as a local action connected with regions (lines for simplicity) of zero modulus of the Weyl field. The self-interaction term $\mathrm{tr} A_m I_m$ is finite and makes contribution to kinetic energy contributing to the mass of a particle. The term $\mathrm{tr} A_m I_n$ ($m \neq n$) describes an interaction between regions of $|\varphi| = 0$. (The wrong sign is connected with the fact that we choose for connection $A$ anti-Hermitian operators.) To complete the classical picture we should forget about the difference between the new and the old metric relations. This can be justified when the places of the Weyl field variations are well localised. It is worth noting that the Weyl field would provide Poincare stresses introduced with an idea to stabilize elementary particles.

The example (8)-(12) is a simple illustration of how classical physics can be realised due to non-trivial topology of a space-time provided by the Weyl field in conformal invariant theory. In a more complicated scenario, a separate dilaton field could be added to the theory

(or the modulus of Weyl field may be regarded as a dilaton field), see references in [14]. The presence of an additional Higgs field (that would generate masses of particles and should have much higher expectation value than the Weyl/dilaton fields) may be unnecessary as masses of particles could be generated by self-interaction term i.e., by the energy of the field produced by the particle - the point of view shared by Poincare. It is worth noting that the process of particle-antiparticle annihilation provides a strong indication in favour of this hypothesis. By doing Lorentz transformation of a particle-like solution where the Weyl field equals to zero along the world line ($t$, 0) and using the Lorenz invariance of the theory, we can easily check that the relativistic energy-momentum relation holds in both global (as the integrals over the space) and local (as the property of the particle world-lines). The mass of the particle-like solution is proportional to the total energy of the system. Different masses of different particles could correspond to different structures of nodes of the Weyl field.

The fermion part of the action has some subtlety. The conform invariant action can be written as

$$S = S_\psi + S_A = \int \left[ \overline{\psi} * \hat{\theta}_s \wedge D\psi + D\overline{\psi} \wedge *\hat{\theta}_s \psi \right] + \frac{1}{4} \int \mathrm{tr} F \wedge *F \qquad (13)$$

with the current

$$J = -i\overline{\psi} * \hat{\theta}_s \psi . \qquad (14)$$

When space-time has nontrivial topology produced by regions of zeros of the Weyl field, non-trivial associated fibre bundles of spinor fields are possible. Here, the geometry trick can be used to make the density of the charge being constant in the space (in agreement with Dirac "sea of electrons"). As a result, the energy of a fermion will be produced by the energy of self-interacting fields plus small additional energy connected with spin degree of freedom (which corresponds to deviation from the uniformly charged space.) The field contribution to the total energy in this case can still be written in a simple form (12).

**IV. Discussion of Weyl field properties.**

Here we briefly discuss some general properties of the Weyl field in a conformal invariant theory. First, we note that the Weyl field is bosonic field, which follows from the fact that (1) is a scalar with respect to Lorentz transformations. It is worth noting that normally it is gauge fields that are bosonic (which follows from the Lorentz invariance of $\hat{T}_a A^a_\mu dx^\mu$ and the fact that $dx^\mu$ is transformed by bosonic (1/2,1/2) Lorentz representation [13]), while elementary fields of matter are fermionic. Therefore, if the Weyl field does exist it might be a composite field (in an analogy with a superconductive condensate). Second, the conformal invariant action (8) may naturally provide Einstein gravity (the term $\frac{|\varphi|^2}{2} R^{ab} \wedge \eta_{ab}$) as well as the cosmological constant (the volume term $\frac{\Lambda}{4}|\varphi|^4 d\eta$), which was noticed by many authors [11]. Combined together, these two terms could yield spontaneous symmetry breaking [14] and a nonzero mass for an elementary particle (that is not a gauge particle) in the presence of non-trivial curvature. Third, the curvature of space time is connected to the presence of matter and hence to density of regions with zero modulus of the Weyl field. The particle creation and annihilation and their dynamics is an evolution of zero-Weyl-field regions.

**V. Conclusions.**

We showed that the canonical distance form factorized by the Weyl field suggests a way to solve the problem of particles existence in gauge theories. In this approach, elementary particles represent non-trivial associated fibre bundles realised around regions of space-time where the modulus of the Weyl field is zero and the metric relations are not defined. We discussed how a conformal invariant theory of the Weyl field provides an apology for framework of classical physics.